\documentclass[aps,prb,preprint,showpacs]{revtex4}

\usepackage{epsfig}

\newcommand{\beq}{\begin{equation}}
\newcommand{\eeq}{\end{equation}}
\newcommand{\beqarray}{\begin{eqnarray}}
\newcommand{\eeqarray}{\end{eqnarray}}
\newcommand{\sgn}[2][]{\ensuremath{\text{sgn}_{#1}(#2)}} % sign function
\newcommand{\Kf}[1][\nu\sigma]{\ensuremath{\hat{F}_{#1}}} % Klein factor
\newcommand{\com}[2]{\ensuremath{\left[#1,#2\right]_{-}}} % commutator
\newcommand{\tfrac}{\textstyle\frac}
\newcommand{\half}{\ensuremath{\tfrac{1}{2}}}
\newcommand{\vF}{\ensuremath{v_{F}}} % Fermi velocity
\newcommand{\kF}{\ensuremath{k_{F}}} % Fermi vector
\newcommand{\Hc}{\ensuremath{\mbox{H.c.}}} % Hermitian conjugate
\newcommand{\Ham}[1][]{\ensuremath{{\cal{H}}_{\text{\tiny{#1}}}}} % Hamiltonian
\newcommand{\CO}[2][\alpha]{\ensuremath{\Lambda^{#2}_{#1}(k)}} % cut-off function
\newcommand{\notag}{\nonumber}
\newcommand{\eq}[1]{Eq.~(\ref{#1})} % Eq. label
\newcommand{\fig}[1]{Fig.~(\ref{#1})} % Fig. label

\begin{document}

\title{Electronic polarons in an extended Falicov-Kimball model} 
\author{P. M. R. Brydon${}^{\rm 1}$, M. Gul\'{a}csi${}^{\rm 1}$ and
A. R. Bishop${}^{\rm 2}$} 
\affiliation{
${}^{\rm 1}$ 
Department of Theoretical Physics, 
Institute of Advanced Studies, 
The Australian National University, Canberra, ACT 0200, Australia \\
${}^{\rm 2}$ 
Theoretical Division, Los Alamos National Laboratory, 
Los Alamos, NM 87545, U.S.A.}

\date{\today}

\begin{abstract}
We examine the one-dimensional spinless Falicov-Kimball model extended
by a hybridization potential between the localized and itinerant
electron states. Below half-filling we find a crossover from a
mixed-valence metal to an integer-valence phase separated state with
increasing on-site Coulomb repulsion. This crossover regime is
characterized by local competition between the strong- and
weak-coupling behaviour, manifested by the formation of an electronic
polaron liquid. We identify this intermediate-coupling regime as a
charge-analogy of the Griffiths phase; a phase diagram is presented
and discussed in detail.  
\end{abstract}

\maketitle

The Falicov-Kimball model (FKM) has found widespread use as a model of
the charge physics in mixed-valence (MV) materials. Although the FKM's
description of a band of itinerant conduction ($c$-)electrons
interacting via a repulsive contact potential $G$ with an underlying
lattice of localized ($f$-)electrons retains its
relevance,~\cite{FKMoriginal} the great evolution of our understanding 
of MV systems has required modification of the model's original
form.~\cite{LRP81} The most popular such modification is the assumption
of some overlap between the $c$- and $f$-wavefunctions: this yields
the extended or quantum Falicov-Kimball model (QFKM).

The ground state of the QFKM has for many years been a matter of
controversy, with two contradictory pictures common in the
literature. The homogeneous solution~\cite{homogeneous} assumes a
simple renormalization of the 
$G=0$ band structure by the Coulomb interaction; at mean-field level,
this has been used to deduce the existence of a novel electronic
ferroelectric MV state.~\cite{POS96} On the other hand, the
relationship of the single-impurity QFKM to the Kondo
impurity~\cite{SchlottmannKondo} suggests distinct strong- and
weak-coupling regimes in the lattice
model.~\cite{SchlottmannRG,epolarons} 

The latter scenario leads to
interesting properties of the intermediate-coupling regime,
specifically the formation of electronic polarons.~\cite{epolarons,T54}
This electronic polaron model has recently been used to explain the
transport properties of certain Heusler and Ce alloys;~\cite{epolexp} more
intriguingly, such polaronic structures play a pivotal role in
the theory of the exotic ``valence-fluctuating'' superconductivity in
CeCu${}_{2}$Ge${}_{2}$ and CeCu${}_{2}$Si${}_{2}$.~\cite{vfsuper}  It
is therefore of considerable interest to study the formation of
electronic polarons in the QFKM. 

In this letter we address this question by a non-perturbative study of
the one-dimensional (1D) QFKM below half-filling. We derive an
effective Hamiltonian for the occupation of the $f$-orbitals which
gives a complete description of the physics. This allows us to exactly
study the intermediate-coupling regime, where we find local competition
between the strong- and weak-coupling phases. This produces a charge
analogy of the Griffiths phase, which may be visualized as a gas of
electronic polarons moving in a MV background. A schematic phase
diagram is presented and discussed in detail.

The 1D QFKM for spinless fermions has the Hamiltonian  
\beq
\Ham[QFKM] = -t\sum_{\langle{i,j}\rangle}c^{\dagger}_{i}c_{j} +
\epsilon_{f}\sum_{j}n^{f}_{j} + G\sum_{j}n^{c}_{j}n^{f}_{j} +
V\sum_{j}\left\{c^{\dagger}_{j}f_{j} + \Hc\right\}  \label{eq:QFKM}
\eeq
Some overlap between the $c$- and $f$-electron wavefunctions is
assumed hence the on-site hybridization term $V\ll{t}$. The
concentration of electrons is fixed at $n =
(1/N)\sum_{j}\left\{n^{f}_{j}+n^{c}_{j}\right\}$ 
where $N$ is the number of sites. In general we have $n<1$ due to
the presence of doped impurities. We assume a MV state at $G=0$, requiring
the Fermi level to be pinned at $-2t<\epsilon_{f}<-2t\cos(n\pi)$ in
the noninteracting system.~\cite{LRP81}

Bosonization is an asymptotically exact method, rigorously describing
the long-wavelength collective behaviour
of the itinerant electrons;~\cite{vDS98} the short-range behaviour remains
fermionic, however, and a cut-off $\alpha$ on the wavelength of the
bosonic density fluctuations is therefore imposed.~\cite{G04} In
simple cases such as the Hubbard model, $\alpha$ can be taken as
vanishing. This is not true for systems such as the QFKM where the
interactions with the localized orbitals determine the lattice
constant as the minimum physical length scale. We are thus required to
assume finite $\alpha>a$: in the QFKM $\alpha$ parameterizes the
electron delocalization length. 

A finite wavelength cut-off may be easily included into the standard
bosonization formalism.~\cite{G04} In particular, for the
linearized fermionic fields about the Fermi points the well-known
Mandelstram representation still holds
\beq
c_{\nu{j}} =
\sqrt{\frac{{A}a}{\alpha}}\Kf[\nu]\exp\left(-i\nu\left[\phi(x_j) -
\nu\theta(x_j)\right]\right) \label{eq:B:b_identity}
\eeq
where $\phi$ and $\theta$ are the dual Bose fields, and the subscript
$\nu=L(-)$, $R(+)$ as subscript (otherwise) for the left- and
right-moving fermions respectively. The Bose fields are defined in
terms of particle-hole excitations about the two Fermi points. For a
system of length $L\gg{a}$ we have
\beqarray
\phi(x_{j}) &=&
-i\sum_{\nu}\sum_{k\neq0}\frac{\pi}{kL}\rho_{\nu}(k)\CO{}{e}^{ikx_j}
\label{eq:Bfieldphi} \\
\theta(x_{j}) &=&
-i\sum_{\nu}\sum_{k\neq0}\nu\frac{\pi}{kL}\rho_{\nu}(k)\CO{}{e}^{ikx_j} \label{eq:Bfieldtheta}
\eeqarray
where $\rho_{\nu}(k)=\sum_{k'}c^{\dagger}_{\nu,k'-k}c_{\nu,k'}$ are
the particle-hole density operators. The wavelength cut-off is
enforced in~\eq{eq:Bfieldphi} and~\eq{eq:Bfieldtheta} by the function
$\CO{}$: this function satisfies the conditions $\CO{}\approx1$ for
$|k|<\alpha^{-1}$ and $\CO{}\approx0$ otherwise. The numerical
constant $A$ in~\eq{eq:B:b_identity} is determined by the functional
form of $\CO{}$. $\Kf[\nu]$ is the so-called Klein factor.~\cite{G04,vDS98}

\eq{eq:B:b_identity} correctly reproduces the long-wavelength
($\gg{a}$) fermionic anticommutators and correlation functions. Since
the Bose fields cannot resolve separations less than $\alpha$, however,
the Mandelstram identity breaks down at short distances
$\sim{a}$. This is reflected in the Bose field commutators, which are
``smeared'' by the cut-off function. For example, assuming exponential
cut-off $\CO{}=\exp(-\alpha|k|/2)$ we have
$\com{\phi(x_{j})}{\theta(0)} = -i\arctan({x_j}/\alpha)$ and 
$\com{\partial_{x}\phi(x_{j})}{\theta(0)}=-i\alpha[\alpha^2+(x_j)^2]^{-1}$.
In the limit $\alpha\rightarrow0$ we recover the Luttinger model 
forms of a sign and Dirac delta function respectively.~\cite{vDS98}   

Substituting~\eq{eq:B:b_identity} into~\eq{eq:QFKM} and applying
standard field-theory techniques, it is a simple matter to obtain the
bosonized form of the QFKM Hamiltonian. For insight into the
physics, however, we must rotate the Hilbert space to couple the $c$-
and $f$-electron orbitals. To accomplish this, we apply a lattice
generalization of the shift transformation used by Schotte and Schotte
in the X-ray edge problem, $\hat{U} =
\exp\left\{i\frac{Ga}{\pi\vF}\sum_{j}(n^{f}_{j}-\half)\theta(x_j)\right\}$.~\cite{SS69}
The Fermi velocity is defined $\vF=2ta\sin(\kF{a})$ where
$\kF=n^{c}_{0}\pi/a$, with $n^{c}_{0}$ the noninteracting $c$-electron
concentration. 

In the transformed Hamiltonian we combine the
$f$-electron operators and the Klein factors 
into pseudospins using a generalized Jordan-Wigner transformation
$\tau^{z}_{j} = n^{f}_{j}-\half,
\tau^{+}_{j}=f^{\dagger}_{j}\Kf[\nu]e^{-i\nu\pi{x_{j}}/2a}{\exp\left({-i\frac{\nu\pi}{2}\sum_{j'}\sgn{x_{j'}-x_j}\left(n^{f}_{j'\sigma}-\half\right)}\right)}$. 
Note the relationship of $\tau^{z}_{j}$ to the occupation of the
$f$-orbital at site $j$. We then write the transformed Hamiltonian in
its final form
\beqarray
{\hat{U}}^{\dagger}\Ham[FKM]{\hat{U}}&=&\frac{\vF{a}}{2\pi}\sum_{j}\left\{\left(\partial_{x}\phi(x_j)\right)^{2}
+ \left(\partial_{x}\theta(x_j)\right)^{2}\right\} +
{G}\left(n^{c}_{0}-\half\right)\sum_{j}\tau^{z}_{j} \notag\\ 
& &-{\frac{G^2a^2}{2\pi^2\vF}\sum_{j,j'}\tau^{z}_{j}\{i\com{\partial_{x}\phi(x_{j})}{\theta(x_{j'})}\}\tau^{z}_{j'}}
\notag \\
&&-\frac{2GAa}{\alpha}\sum_{j}\tau^{z}_{j}\cos\left(2\left\{\phi(x_j)-
{\cal{K}}_{\alpha}(x_j)
-\left[\kF+\frac{\pi}{2a}\right]x_j\right\}\right)\notag \\
&&
+2V\sqrt{\frac{Aa}{\alpha}}\sum_{j}\left\{{\tau}^{+}_{j}e^{i\left(1-\frac{Ga}{\pi\vF}\right)\theta(x_j)}\cos\left(\phi(x_j)-{\cal{K}}_{\alpha}(x_j)
- \left[\kF + \frac{\pi}{2a}\right]x_j\right) + \Hc\right\}
\label{eq:EPH:CTHam} 
\eeqarray
where ${\cal{K}}_{\alpha}(x_j) =
{\cal{S}}_{\alpha}(x_j)+{\cal{L}}_{\alpha}(x_j)$,
${\cal{S}}_{\alpha}(x_j) =
\sum_{n=1}^{\infty}[\{i\com{\theta(x_{j+n})}{\phi(x_{j})}\}-\frac{\pi}{2}]\left(\tau^{z}_{j+n}-\tau^{z}_{j-n+1}\right)$ 
and  
${\cal{L}}_{\alpha}(x_j) =
\left(\frac{Ga}{\pi\vF}-1\right)\sum_{n=1}^{\infty}\{i\com{\theta(x_{j+n})}{\phi(x_{j})}\}\left(\tau^{z}_{j+n}-\tau^{z}_{j-n}\right)$.
We have kept all terms produced by the canonical transform
in~\eq{eq:EPH:CTHam}. 

The three terms in~\eq{eq:EPH:CTHam} arising from the Coulomb interaction
are of special note. The first term represents a renormalization of
the $f$-level energy and drives a valence transition with increasing
$G$. For $0\leq{n^c_0}<0.5$ ($0.5<{n^c_0}\leq1$) the negative (positive) 
sign of this energy-shift implies the emptying of the $c$-electron
($f$-electron) band so that all electrons have an unambiguously
$f$-electron ($c$-electron) character. The case 
$0\leq{n^{c}_{0}}<0.5$ is particularly interesting as here the Ising
interaction term orders the available $f$-electrons into a single contiguous
block, the well-known segregated (SEG) phase
of the $V=0$ limit.~\cite{L92} The origin of this interaction is the finite
spread $\alpha>a$ of the $c$-electron wavefunctions: to minimize the
interorbital Coulomb repulsion this favours empty underlying
$f$-orbitals. The last term originates from the backscattering of the
$c$-electrons off the localized 
orbitals. In the absence of the hybridization, this will order the
available $f$-electrons into crystalline phases via a Peierls-like
mechanism.~\cite{FF90} The addition of the hybridization, however,
replaces these phases by a MV state; in the following
analysis we therefore neglect the backscattering. Including this
effect does not qualitatively alter our conclusions.~\cite{BG05}

The phases of the QFKM are most usefully classified in terms of the
occupation of the localized orbitals. We therefore derive an effective
Hamiltonian \emph{only} for the $f$-occupation (i.e. the
$\tau$-pseudospins). This is straightforwardly achieved by replacing
the Bose fields in~\eq{eq:EPH:CTHam} with suitably chosen expectation
values. At weak-coupling, Schlottmann's renormalization group study
reveals that the system flows to the $G=0$ fixed
point;~\cite{SchlottmannRG} at strong-coupling, where the system is in
the SEG phase, exact diagonalization calculations by
Farka\v{s}ovsk\'{y} in the $V=0$ limit indicate that the $c$-electrons
are at their noninteracting fixed point.~\cite{F03} On the basis of
these studies it is natural to take $\langle\phi(x_j)\rangle =
\langle\theta(x_j)\rangle = 0$ across the phase diagram. We hence
obtain the effective Hamiltonian 
\beq
\Ham[eff] =
-{\cal{J}}\sum_{j}\tau^{z}_{j}\tau^{z}_{j+1}
+G\left(n^{c}_{0}-\half\right)\sum_{j}\tau^{z}_{j} +
4V\sqrt{\frac{Aa}{\alpha}}\sum_{j}{\tau}^{x}_{j}
\cos\left({\cal{K}}_{\alpha}(x_j)+ \left[\kF + \frac{\pi}{2a}\right]x_j\right)
\label{eq:EPH:effHam} 
\eeq
We have approximated the segregating interaction by its nearest-neighbour
form where
${\cal{J}}=\frac{G^2a^2}{\pi\vF}\frac{1}{L}\sum_{k}\CO{}e^{ika}$.

$\Ham[eff]$ is instantly recognizable as a transverse field Ising
model. For large $G$, we recover the SEG phase as the Ising term
dominates the Hamiltonian. 
In the limit of vanishing Coulomb repulsion, however, the
site-dependent transverse field
$h^{x}_{j}=4V\sqrt{\frac{Aa}{\alpha}}\cos({\cal{K}}_{\alpha}(x_j)+[\kF+\frac{\pi}{2a}]x_j)$ 
dominates~\eq{eq:EPH:effHam}. The   
pseudospins are therefore ordered along the $x$-axis with
$\langle\tau^{x}_{j}\rangle\approx\half\sgn{h^{x}_{j}}$. Translated 
back into the $f$-occupation `language' this implies that 
$\langle{f^{\dagger}_{j}c_{j}+\Hc}\rangle\neq0$, with the occupation
of each $f$-site being given by
$\langle{n^{f}_{j}}\rangle\approx{n^{f}_{0}}$  
where $n^{f}_{0}$ is the $f$-electron concentration for $G=0$.
This clearly corresponds to a MV state.

The MV and SEG phases are characterized by quite contradictory
behaviours of the $f$-electrons: in the former the localized orbitals
are in a superposition of their different valence states, while in the
latter there is a phase separation between regions with different
integer ionic valence. At coupling strengths intermediate between the
MV and SEG phases, therefore, we expect to observe some competition
between these two distinct charge physics.
For a vanishing longitudinal field, \eq{eq:EPH:effHam} displays an
order-disorder transition; the longitudinal field however lifts the
system from criticality and we instead find a crossover between
the ordered and disordered phases.~\cite{UJP80} To understand the
details of the crossover regime, we must examine the site-dependence of
$h^{x}_{j}$; specifically, we study the operator
${\cal{K}}_{\alpha}(x_j)$, which we replace by its expectation
value.  
 
The string operator ${\cal{K}}_{\alpha}(x_j)$ consists of two
contributions, one measuring the short-range [${\cal{S}}_{\alpha}(x_j)$] and
the other the long-range [${\cal{L}}_{\alpha}(x_j)$] ordering of the
pseudospins. In 
the SEG phase, the short-range correlations of the pseudospins are very
strong and so ${\cal{S}}_{\alpha}(x_j)$ vanishes; the 
importance of disorder in the crossover regime, however, means that
here ${\cal{S}}_{\alpha}(x_j)$ has a random variation. In
contrast, the long-range ordering of the pseudospins in the SEG
phase persists deep into the crossover regime. ${\cal{L}}_{\alpha}(x_j)$ is
only non-zero for $0\leq{n^{c}_{0}}<0.5$ where it takes a maximum at
the edge of  the $f$-electron block, increasing linearly as this point is
approached from either side: for the boundary of the
$f$-block at $x_{0}$ we find 
${\cal{L}}_{\alpha}(x_j)\sim\left(\frac{Ga}{\pi\vF}-1\right)|j|$.
The linear variation of ${\cal{L}}_{\alpha}(x_j)$ implies a
quasiperidoic transverse field in the effective
Hamiltonian~\eq{eq:EPH:effHam}. The periodicity of this field is
in general incommensurate with the lattice: numerical~\cite{Satija} and
analytical~\cite{JK93} studies have revealed the physics to be
qualitatively 
identical to a random variation of transverse field. Fisher has
used a real-space renormalization group method to exmaine in
detail this random transverse field Ising model.~\cite{F95} To
utilise his detailed results we replace the transverse field in
\eq{eq:EPH:effHam} by a field ${\widetilde{h}}^{x}_{j}$ which varies
randomly with $j$, the values being drawn from a cosine distribution
$\rho(\widetilde{h})d\widetilde{h}=(C\pi)^{-1}\sqrt{1-(\widetilde{h}/C)^2}$
where $C=4V\sqrt{Aa/\alpha}$. 

\begin{figure}[p]
\includegraphics{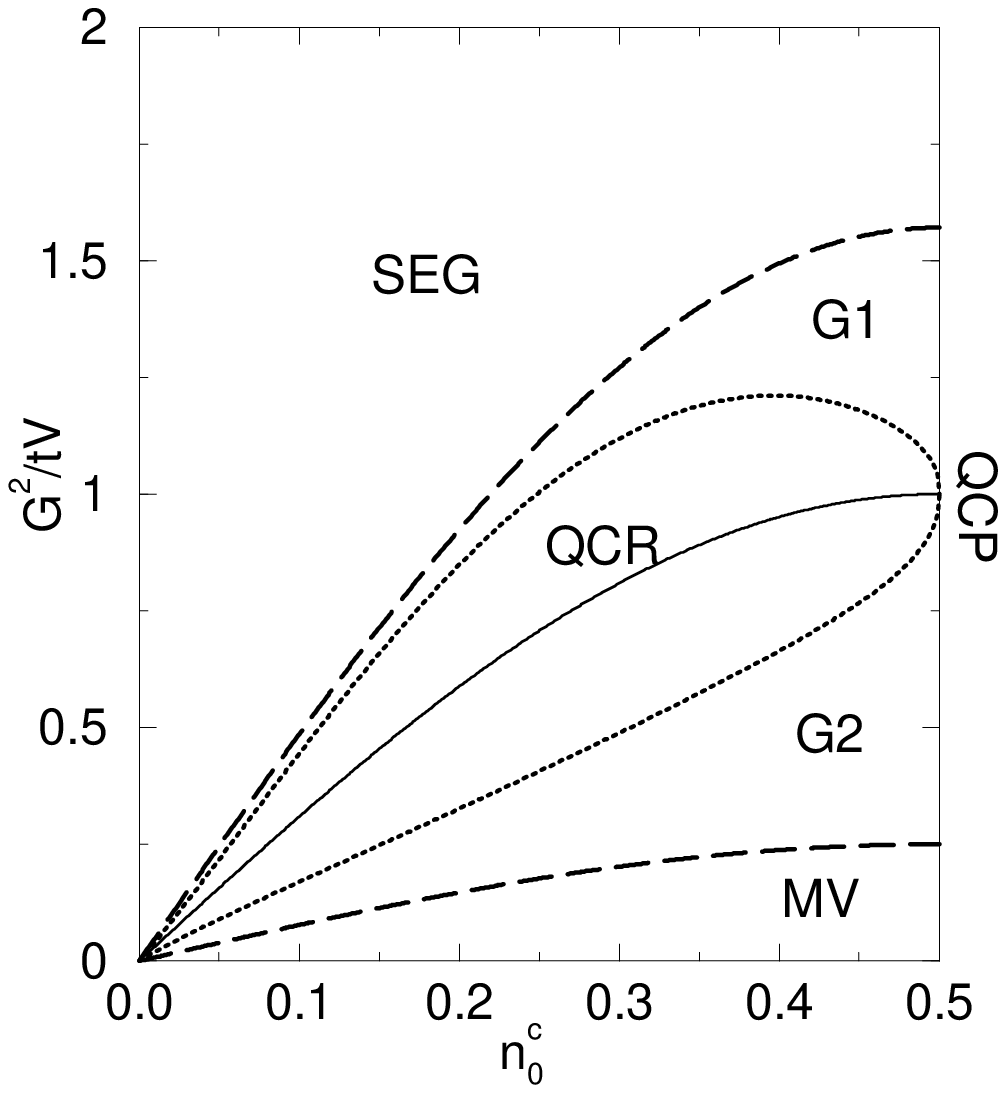} 
\caption{\label{fig:PD}Ground-state phase diagram for the QFKM below
half-filling. The quantum critical regime (QCR) is found on either
side of the solid line $\delta=0$. Along the line $n^{c}_{0}=0.5$ the QCR
is replaced by a quantum critical point (QCP). The dotted lines denote the
approximate ``boundary'' of the QCR with the Griffiths phases G1 and G2. 
The dashed lines separate these Griffiths phases from the segregated
(SEG) and mixed-valence (MV) phases. Electronic polarons are present
in both Griffiths phases and the QCR.}
\end{figure}

Following Fisher we find it convenient to define the dimensionless
parameter $\delta=\log(G_{c}/G)$ where $G^2_{c} =
kt\sin(n^{c}_0\pi)V$. Assuming exponential cut-off, the
proportionality constant takes the form
$k=8\sqrt{\frac{2\pi{a}}{\alpha}}\left[\frac{\alpha}{a}+\frac{a}{\alpha}\right]$.
Although the dependence upon $\alpha$ obviously restricts the predictive
powers of our analysis, it does not however prevent us from drawing a
schematic phase diagram: for this purpose it is acceptable to choose
$\alpha=\text{constant}$. Without loss of generality we therefore set
$k$ equal to unity, being equivalent to renormalizing the Coulomb
interaction $G\rightarrow{G/\sqrt{k}}$. In the QFKM's phase diagram
presented in \fig{fig:PD} we plot the ratio $G^2/tV$ as a function of
$n^{c}_{0}$. We note that the phase diagram for
$0.5<{n^{c}_{0}}\leq1$ can be obtained by reflecting \fig{fig:PD}
along the line $n^{c}_{0}=0.5$. The only significant difference to 
\fig{fig:PD} is that the SEG phase is replaced by the ``empty''
state ($n^{f}=0$). We obtain the $V\rightarrow0$ weak-coupling
crystalline phases in the phase diagram on the inclusion of the 
backscattering corrections in~\eq{eq:EPH:CTHam}. Physically, since
a finite hybridization is always present, a MV state will be
realized instead of these phases, as we have observed above. 

Starting in the segregated phase ($\delta\ll0$) we lower $G$. Below
the value $G=\sqrt{\pi/2}G_{c}$, the Ising coupling in
\eq{eq:EPH:effHam} is no longer greater than the transverse field
\emph{everywhere} on the lattice. The rare regions where the
transverse field term is strongest breaks the single $f$-electron
block up into randomly distributed large clusters separated by regions
of the MV phase: this local competition between the MV and SEG
phases may be regarded as a charge analogy of the Griffiths
phase [G1 in~\fig{fig:PD}].~\cite{F95}
Although such a state has been inferred by numerical work, this
is the first analytic indication of its existence.~\cite{MD01} 
The system still, however, retains recognizable features of the SEG phase:
the mean $f-f$ correlations
$\overline{\langle{n^{f}_{j}n^{f}_{j+x}}\rangle}$ decay exponentially
towards $(n^{f})^2>(n^{f}_{0})^2$ with a correlation length
$\xi\sim{e^{-2\delta^2}}$, the non-universal power-law dependence
reflecting the continued presence of the infinite $f$-cluster
throughout G1.

As $G$ is lowered the Griffiths phase G1 gives way to the quantum
critical regime (QCR). The QCR is dominated by the physics of the
critical point: at every length scale we find the coexistence of equal
regions of mixed- and integer-valence. Due to the 
$f$-level renormalization we do not find algebraic decay of
correlation functions, but rather exponential attenuation with a
correlation length of $\xi\sim1+{\cal{O}}(\delta^2)$. For
$n^{c}_{0}=0.5$ the $f$-level shifting vanishes and so the QCR shrinks
down to a single quantum critical point (QCP) characterized by
dynamical critical exponent $z=\infty$ at
$G=G_{c}(n^{c}_{0}=0.5)=\sqrt{tV}$. Note that the precise position of
the QCP is dependent upon the choice of $\alpha$.~\cite{BG05}

Further decreasing $G$ below the QCR, we reach a second Griffiths phase (G2)
with opposite character to G1: here the hybridization term dominates
the segregating interaction over most of the lattice. The SEG phase
almost completely disappears, with only rare clusters of valence-ordered
localized orbitals embedded in an MV background remaining of this
state. The mean correlation functions decay exponentially, although
less rapidly than in the G1, with $\xi\sim(1+\delta^{-2})^{-1}$. Since the
minimum magnitude that the transverse field in $\Ham[eff]$ assumes is
$\min{|{\widetilde{h}}^{x}_j|}=0$, the weakly disordered phase should
be present for any $G>0$; for $\delta\gg1$ the integer-valence
clusters are however extremely rare and the last two terms of
$\Ham[eff]$ dictate the physics. As such, we can justifiably identify 
the very low-$G$ behaviour as typical of a MV state [\fig{fig:PD}]. 

The characteristic feature of the Griffiths phases and the QCR is the
local competition between the MV and SEG states. This manifests itself as
the co-existing clusters of $f$-orbitals with mixed- and
integer-valence. The integer-valence clusters are of greatest 
interest: these correspond to the dressing of a $c$-electron by a
``cloud'' of near-empty $f$-orbitals. This coupling of the $c$- and
$f$-electron densities originates from the forward-scattering, and
constitutes an electronic polaron. Electronic
polarons form only at intermediate-coupling: at weak-coupling the
screening clouds are heavily suppressed by the resonant (MV) scattering
with the $f$-orbitals whereas at strong coupling the clouds merge to
form the SEG phase. Since the bosonization method treats the
forward-scattering exactly,
our discovery of electronic polarons within a
crossover regime between mixed- and integer-valence phases
\emph{rigorously} confirms the scenario proposed by Liu and
Ho.~\cite{epolarons} 
  
Note that in the Griffiths phases which flank the sides of the QCR the
length scale $\xi$ differs from the conduction 
electron mean free path which gives rise to competing time scales:
slow motion of the electronic polarons and fast motion of the
conduction electrons. The different dynamics of the two types of
particles provides a close analogy to a two fluid scenario. Since the
polarons are randomly distributed across the lattice, these
states can be viewed as intrinsic inhomogeneities involving charge
fluctuations and short-range charge correlations. This resembles very
closely the spin polaron liquid found in the 1D Kondo lattice
model.~\cite{McCJRG02}  

Summarizing our results, we have derived an effective model for the
occupation of the localized orbitals in the 1D QFKM. This model
predicts a crossover from a MV state at
weak $c-f$ Coulomb repulsion $G$ to the integer-valence SEG state with
increasing coupling strength. At intermediate values of the coupling the
system exhibits an electronic Griffiths phase. The effective
Hamiltonian gives a detailed accounting of the $f$-electron physics
across the phase diagram, confirming the importance of electronic
polarons in the QFKM. An extensive discussion of this system is to
follow in a later paper\cite{BGinprep}.


\begin{thebibliography}{99}

\bibitem{FKMoriginal}L. M. Falicov and J. C. Kimball,
Phys. Rev. Lett. {\bf{22}}, 997 (1969).

\bibitem{LRP81}J. M. Lawrence, P. S. Riseborough and R. D. Parks,
Rep. Prog. Phys. {\bf{44}}, 1 (1981).

\bibitem{homogeneous}H. J. Leder, Solid State Comm. {\bf{27}}, 579
  (1979); W. Hanke and J. E. Hirsch, Phys. Rev. B {\bf{25}}, 6748 (1982). 

\bibitem{POS96}T. Portengen, Th. \"{O}streich and L. J. Sham,
  Phys. Rev. B {\bf{54}}, 17452 (1996). 

\bibitem{SchlottmannKondo}P. Schlottmann, Solid State Comm. {\bf{31}}, 885
(1979); Phys. Rev. B {\bf{22}}, 622 (1980).

\bibitem{SchlottmannRG}P. Schlottmann, Phys. Rev. B {\bf{22}}, 613 (1980).

\bibitem{epolarons}S. H. Liu and K.-M. Ho, Phys. Rev. B {\bf{30}},
  3039 (1984); S. H. Liu, Phys. Rev. Lett. {\bf{58}}, 2706 (1987).

\bibitem{T54}Y. Toyozawa, Prog. Theor. Phys. {\bf{12}}, 421 (1954).

\bibitem{epolexp}A. \'{S}lebarski {\it{et al.}}, Phys. Rev. B
  {\bf{69}}, 155118 (2004); N. E. Sluchanko {\it{et al.}},
  cond-mat/0505386 (unpublished).

\bibitem{vfsuper}Y. Onishi and K. Miyake,
  J. Phys. Soc. Jpn. {\bf{69}}, 3955 (2000) ; A. T. Holmes, D. Jaccard
  and K. Miyake, Phys. Rev. B {\bf{69}}, 024508 (2004). 

\bibitem{G04}M. Gul\'{a}csi, Adv. Phys. {\bf{53}}, 769 (2004).

\bibitem{vDS98}J. von Delft and H. Schoeller, Ann. der Physik
  {\bf{4}}, 223 (1998). 

\bibitem{SS69}K. D. Schotte and U. Schotte, Phys. Rev. {\bf{182}}, 479
(1969). 

\bibitem{F03}P. Farka\v{s}ovsk\'{y}, Intl. J. Mod. Phys. B {\bf{17}},
  4897 (2003).

\bibitem{L92}P. Lemberger, J. Phys. A: Math. Gen. {\bf{25}}, 715
  (1992).

\bibitem{FF90}J. K. Freericks and L. M. Falicov, Phys. Rev. B
  {\bf{41}}, 2163 (1990). 

\bibitem{BG05}P. M. R. Brydon and M. Gul\'{a}csi, cond-mat/0506218
  (unpublished).

\bibitem{UJP80}K. Uzelac, R. Jullien and P. Pfeuty, Phys. Rev. B
  {\bf{22}}, 436 (1980). 

\bibitem{Satija}I. I. Satija and M. M. Doria, Phys. Rev. B {\bf{39}},
  9757 (1989); I. I. Satija, Phys. Rev. B {\bf{41}}, 7235 (1990). 

\bibitem{JK93}S. Jitormirskaya and A. Klein, J. Stat. Phys. {\bf{73}},
  319 (1993).

\bibitem{F95}D. S. Fisher, Phys. Rev. B {\bf{51}}, 6411 (1995).

\bibitem{MD01}E. Miranda and V. Dobrosavljevi\'{c},
  Phys. Rev. Lett. {\bf{86}}, 264 (2001). 

\bibitem{McCJRG02}I. P. McCulloch, A. Juozapavicius, A. Rosengren and
  M. Gul\'{a}csi, Phys. Rev. B {\bf{65}}, 052410 (2002).

\bibitem{BGinprep}P. M. R. Brydon and M. Gul\'{a}csi, in preparation.

\end{thebibliography}
\end{document}